%% file: main.tex
\documentclass[aps, pre, twocolumn, superscriptaddress, reprint]{revtex4-1}

\usepackage{graphicx}
\usepackage{bm}
\usepackage{dcolumn}
\usepackage{amssymb,amsmath}
\usepackage[colorlinks=true,allcolors=blue]{hyperref}
\usepackage{color}
\usepackage{siunitx}

\usepackage{lipsum}
\newcommand{\cOut}[1]{}
\usepackage[todonotes={textsize=tiny,textwidth=44pt},truncate=hyphenate,deletedmarkup=xout,addedmarkup=uwave]{changes}
\newcommand{\figref}[2]{\hyperref[#1]{\ref{#1}(#2)}}

\begin{document}
\title{Experimental observation of the curvature-induced asymmetric spin-wave dispersion in hexagonal nanotubes}

\author{Lukas K\"orber}
\email[Corresponding author: ]{l.koerber@hzdr.de}
\affiliation{Helmholtz-Zentrum Dresden~-~Rossendorf, Institute of Ion Beam Physics and Materials Research, Bautzner Landstra{\ss}e 400, 01328 Dresden, Germany}
\affiliation{Fakultät Physik, Technische Universit\"at Dresden, D-01062 Dresden, Germany}

\author{Michael Zimmermann}
\affiliation{Fakultät für Physik, Universit\"at Regensburg, Universit\"atsstra{\ss}e 31, D-93053 Regensburg, Germany}

\author{Sebastian Wintz}
\affiliation{Swiss Light Source, Paul Scherrer Institut, 5232 Villigen PSI, Switzerland}
\affiliation{Max-Planck-Institut für Intelligente Systeme, 70569 Stuttgart, Germany}

\author{Simone Finizio}
\affiliation{Swiss Light Source, Paul Scherrer Institut, 5232 Villigen PSI, Switzerland}

\author{Markus Weigand}
\affiliation{Max-Planck-Institut für Intelligente Systeme, 70569 Stuttgart, Germany}
\affiliation{Helmholtz-Zentrum Berlin, 12489 Berlin, Germany}

\author{J\"org Raabe}
\affiliation{Swiss Light Source, Paul Scherrer Institut, 5232 Villigen PSI, Switzerland}

\author{Jorge A. Ot\'alora}
\affiliation{Departamento de Física, Universidad Católica del Norte, Avenida Angamos 0610, Casilla 1280, Antofagasta, Chile}

\author{Helmut Schultheiss}
\affiliation{Helmholtz-Zentrum Dresden~-~Rossendorf, Institute of Ion Beam Physics and Materials Research, Bautzner Landstra{\ss}e 400, 01328 Dresden, Germany}
\affiliation{Fakultät Physik, Technische Universit\"at Dresden, D-01062 Dresden, Germany}

\author{Elisabeth Josten}
\affiliation{Ernst Ruska-Centre for Microscopy and Spectroscopy with Electrons (ER-C) and Peter Gr\"unberg Institute (PGI), Forschungszentrum J\"ulich, 52425 J\"ulich, Germany}

\author{J\"urgen Lindner}
\affiliation{Helmholtz-Zentrum Dresden~-~Rossendorf, Institute of Ion Beam Physics and Materials Research, Bautzner Landstra{\ss}e 400, 01328 Dresden, Germany}


\author{Christian H. Back}
\affiliation{Physik-Department, Technische Universit\"at M\"unchen, 85748 Garching b. M\"unchen, Germany}
\affiliation{Fakultät für Physik, Universit\"at Regensburg, Universit\"atsstra{\ss}e 31, D-93053 Regensburg, Germany}

\author{Attila K\'akay}
\affiliation{Helmholtz-Zentrum Dresden~-~Rossendorf, Institute of Ion Beam Physics and Materials Research, Bautzner Landstra{\ss}e 400, 01328 Dresden, Germany}

\date{\today}

\begin{abstract}
Theoretical and numerical studies on curved magnetic nano-objects predict numerous exciting effects that can be referred to as magneto-chiral effects, which do not originate from the intrinsic Dzyaloshinskii-Moriya interaction or surface-induced anisotropies. The origin of these chiral effects is the isotropic exchange or the dipole-dipole interaction present in all magnetic materials but renormalized by the curvature. Here, we demonstrate experimentally that curvature induced effects originating from the dipole-dipole interaction are directly observable by measuring spin-wave propagation in magnetic nanotubes with hexagonal cross section using time resolved scanning transmission X-ray microscopy. We show that the dispersion relation is asymmetric upon reversal of the wave vector when the propagation direction is perpendicular to the static magnetization. Therefore counter-propagating spin waves of the same frequency exhibit different wavelenghts. Hexagonal nanotubes have a complex dispersion, resulting from spin-wave modes localised to the flat facets or to the extremely curved regions between the facets. The dispersion relations obtained experimentally and from micromagnetic simulations are in good agreement. 
These results show that spin-wave transport is possible in 3D, and that the dipole-dipole induced magneto-chiral effects are significant.

\end{abstract}

\maketitle

After having been proposed by Bloch in the 1930s~\cite{Bloch30}, the propagation of spin waves (or magnons) -- the elementary excitations in magnetically ordered systems -- has been studied extensively in the past. Because of their peculiar linear and nonlinear characteristics, spin waves
promise great potential in information transport and processing as, \textit{e.g.}, the magnon transistor \cite{Chumak14} and the magnonic diode \cite{Lan15} for multifunctional spin-wave logic applications. Spin waves (including the spatially uniform ferromagnetic resonance precession) have also been proven to be an excellent tool to probe the magnetic characteristics of solids as they are sensitive to spin currents \cite{Vlaminck08,Chauleau2014,Gladii2017},
impurities \cite{Callaway1964,Abeed2019,Mohseni2019}, crystal anisotropies \cite{Gurevich1996} or asymmetric exchange interactions, among others. For example, the presence of an asymmetric interaction such as the Dzyaloshinskii–Moriya interaction (DMI) leads to an asymmetric dispersion and consequently to a nonreciprocal propagation of spin waves, therein \cite{Udvardi2009,Zakeri10,Moon2013,Cort_s_Ortu_o_2013,Kostylev2014,Koerner2015}. Similar non-reciprocal spin-wave propagation is observed in magnetic bilayers \cite{Henry2016,Gallardo2019}.
Therefore, the study of spin-wave propagation is both of a technological as well as a fundamental interest.

While many of the aforementioned effects have been investigated mostly in bulk or in flat thin-film samples, over the last decade, curvature induced effects have been uncovered as a new way to manipulate magnetic equilibria as well as spin dynamics. Numerous theoretical and numerical works have already shown that the surface curvature of magnetic membranes leads to phenomena which are not present in flat geometries of the same material~\cite{Landeros07,Yan12,Otalora13,Yan13,Gaididei14,Pylypovskyi15,Otalora16,Otalora17,Otalora18,Kravchuk18}. As a result of the bending, in conventional soft magnetic materials exotic non-collinear magnetic textures such as skyrmions~\cite{Kravchuk16a} may be stabilized  or magnetisation dynamics can be influenced, leading to left-right symmetry breaking of domain wall motion~\cite{Yan12,Landeros07}, asymmetric spin-wave transport~\cite{Hertel13a,Otalora16} or the emergence of a topological Berry phase~\cite{Dugaev2005}. So far, experimental evidence for curvature-induced effects was shown in~\cite{Dietrich08}, in which the surface curvature resulted in the bending of the domain walls in a flux-closure magnetic structure and more recently in the work by ~\textcite{Volkov2019}. In the latter, it is shown that due to the curvature-induced effect originating from the exchange interaction, domain walls in parabolic permalloy wires are pinned by the curvature gradient and that by measuring the de-pinning of these domain walls the magnitude of the exchange-induced curvilinear effect can be indirectly determined. Similar to the microscopic DMI, curvature-induced effects are associated to the class of magneto-chiral interactions as they often break chiral symmetry in magnetic systems.



The influence of curvature on magnetic equilibria has been shown to be mainly due to a renormalization of the magnetic exchange interaction \cite{Gaididei14,Sheka2019}. Magnetization dynamics, on the other hand, are mostly perturbed by curvature-induced magnetic charges~\cite{Landeros07,Yan12,Yan13}, \textit{i.e.}, by a renormalization of the dipolar interaction. In our recent works \cite{Otalora16, Otalora17}, it was predicted that this leads to an asymmetric dispersion of spin waves in round magnetic nanotubes being in the vortex state.


In this Letter, we demonstrate experimentally that curvature induced effects originating from the dipole-dipole interaction are directly observable by measuring spin-wave transport in magnetic nanotubes with hexagonal cross section in the vortex state, using time-resolved scanning transmission X-ray microscopy (TR-STXM) \cite{VanWaeyenberge2006,Acremann2007,Wintz16}. These results are a direct qualitative confirmation of our theoretical and numerical prediction for round nanotubes~\cite{Otalora16,Otalora17} with the advantage that the hexagonal core/shell structures can be grown with high quality by molecular beam epitaxy.
Moreover, our micromagnetic simulations and TR-STXM experiments show that the mode spectrum of hexagonal tubes, accessible by a simple microwave antenna is much more diverse than that of round tubes and splits into several branches. Besides the modes localized and propagating along the top- and bottom facets of the tube, spin waves propagating on the side facets or localized to the highly curved corners at the junctions of the facets of the hexagonal cross-section are found. At the same time, modes with a similar collective behaviour as the spin waves found in round tubes could also be addressed. The dispersion is predicted to be asymmetric for all branches and for those branches accessible by the TR-STXM frequency range it is found to be the case, which directly demonstrates the presence of curvature-induced magneto-chiral effects.

\begin{figure}[h!]
\begin{center}
\includegraphics[width=\columnwidth]{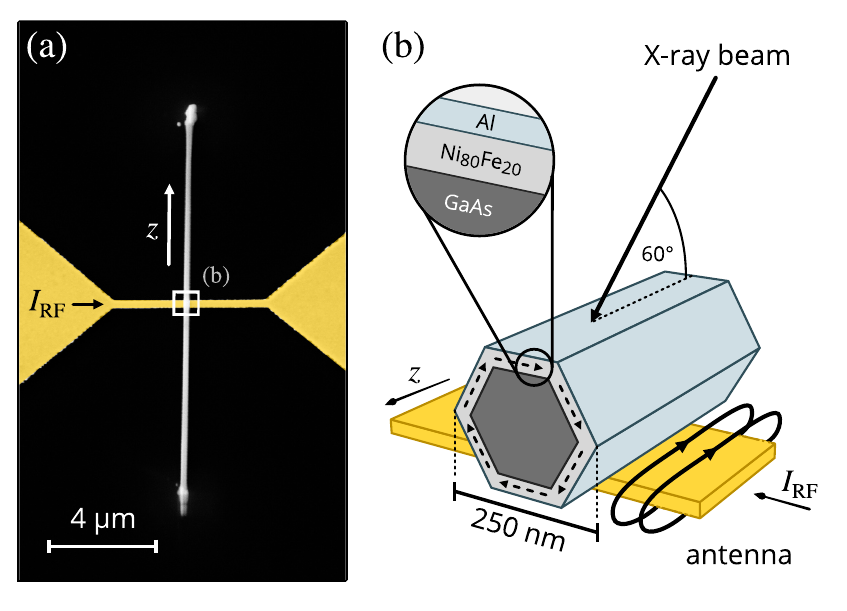}
\caption{\label{fig:fig1}(a) Transmission-electron-microscopy image of a hexagonal permalloy nanotube with \SI{250}{\nano\meter}  outer diameter, \SI{30}{\nano\meter} thickness and \SI{12}{\micro\meter} length on a GaAs wire. The gold strip-like antenna (here, colored for visual purposes) was patterned on a SiN membrane and the nanotube was placed on the top using a focused ion beam (FIB) tool and a micro manipulator. The Oersted field of an rf-current is used to excite spin waves. (b) Cross-sectional sketch of the hexagonal nanotube showing the layer structure. The permalloy layer is directly evaporated on the GaAs wire \textit{in situ} and capped with Al to avoid oxidation. A relative angle of 60 degree was used between the X-ray beam and the symmetry axis of the nanotube. This configuration allows for being sensitive to the in-plane dynamic magnetization of the top- and bottom surfaces of the tube, which is expected to be larger than the out-of-plane dynamic magnetization component.}
\end{center}
\end{figure}

Despite the remarkable achievements on novel approaches for magnetic nano-membranes preparation, such as stretchable \cite{Melzer12}, rolled-up \cite{Schmidt01,Balhorn10} and flexible magnetic membranes, as well as flexible displays \cite{Sugimoto2004}, shapeable nanoelectronics \cite{Ying_2012} and even printable spintronic devices \cite{Karnaushenko12}, to our knowledge, there was no success in the preparation of high-quality round ferromagnetic tubes with sufficiently low damping to allow for experimental studies of spin-wave transport. The closest approach was recently reported by \textcite{Zimmermann18} who succeeded in fabrication of hexagonal ferromagnetic nanotubes with sub-nanometer surface roughness, using a GaAs single crystal rod coated in-situ with Ni$_{80}$Fe$_{20}$ (permalloy) capped with Al for protection. The used incident angle growth conditions during the evaporation process lead to an easy-plane magnetic anisotropy perpendicular to the symmetry axis of the nanotube, resulting in a stable vortex state in the hexagonal nanotube, as confirmed by STXM measurements~\cite{Zimmermann18}. This finally allowed us to experimentally study the curvature-induced effects on spin-wave dynamics in nanotubes. For this purpose, the previously manufactured hexagonal tube with an outer diameter of \SI{250}{\nano\meter}, a thickness of \SI{30}{\nano\meter} and a length of \SI{12}{\micro\meter} was placed on a silicon membrane with a gold antenna underneath to excite spin waves (see Fig.~\ref{fig:fig1}). Spin waves propagating away from the antenna are measured using TR-STXM. The experimental results are complemented by micromagnetic simulations using a custom version of the GPU-accelerated code MuMax$^3$ \cite{Vansteenkiste14} (see Supplemental Materials for details).

\begin{figure}[h!]
\begin{center}
\includegraphics[width=\columnwidth]{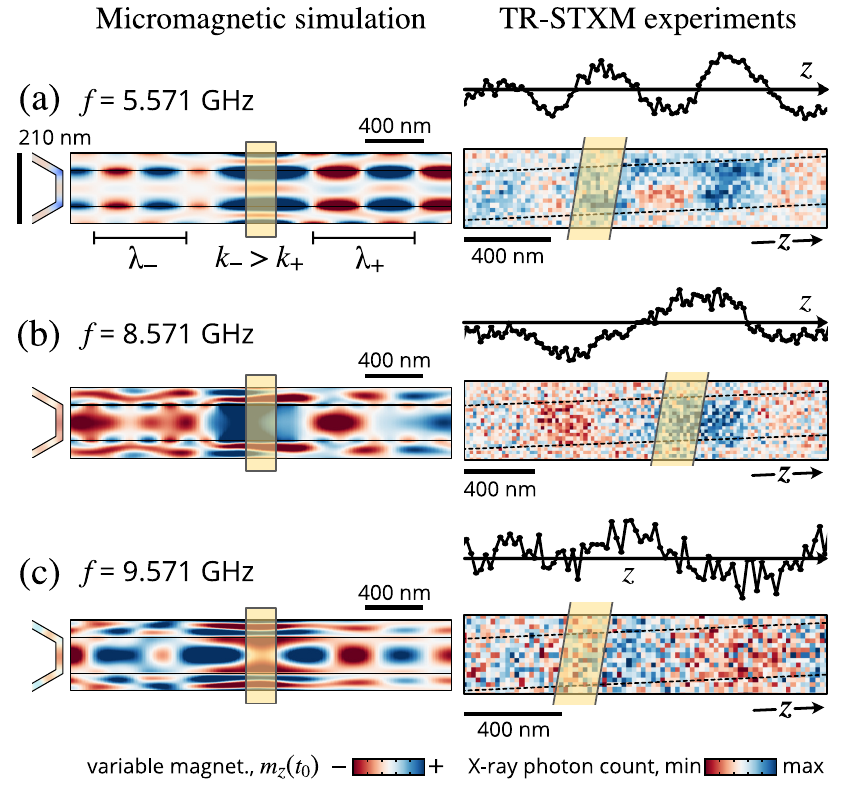}
\caption{\label{fig:fig2}(a)-(c) Numerical (left) and experimental (right) snapshots of the dynamical magnetization at a fixed time for modes propagating in the corners of the hexagonal tube (\SI{5.571}{\giga\hertz}) and for modes mostly propagating on the top and bottom facets (\SI{8.571}{\giga\hertz} and \SI{9.571}{\giga\hertz}). For the numerical profiles, a side view is shown next to the projection of the upper half the nanotube. For better illustration, they have been stretched in the width direction. Above of the experimental profiles, we show averaged linescans along the tube as a visual aid. The position of the antenna is in all cases marked with a translucent gold-colored patch.}
\end{center}
\end{figure}

In the experiments the spin waves were excited at fixed frequencies. In Fig.~\figref{fig:fig2}{a-c}, we show exemplary snapshots of two counter-propagating spin-wave modes at \SI{5.571}{\giga\hertz}, \SI{8.571}{\giga\hertz} and \SI{9.571}{\giga\hertz}, obtained by micromagnetic simulation and TR-STXM, respectively. 
As seen from the profiles, the modes exhibit different localization within the cross section of the hexagonal tube, \textit{e.g.}, there are modes more localized in the corners of the tube (\SI{5.571}{\giga\hertz} in Fig.~\figref{fig:fig2}{a}) or on the facets of the tube (\SI{8.571}{\giga\hertz} and \SI{9.571}{\giga\hertz} in Fig.~\figref{fig:fig2}{b,c}). In this context, we also would like to refer to the experimental movies of these modes, provided in the supplemental material, which show nicely the localization of the modes at \SI{5.571}{\giga\hertz} in the corners of the hexagonal cross section. We observe an intensity asymmetry of the modes at large frequencies. This is a commonly known effect for Damon-Eshbach spin waves (with $\bm{k}\perp\bm{M}_\mathrm{eq}$) that are excited with a strip-line antenna in magnetic thin films \cite{Demidov2009}, and is also present here as our tubes are in the vortex state ($\bm{k}\perp\bm{M}_\mathrm{eq}$). Moreover, in the numerical mode snapshot at \SI{5.571}{\giga\hertz} (Fig.~\figref{fig:fig2}{a}), one can already clearly see a wave-vector asymmetry for the two counter-propagating modes which is the evidence for an asymmetric spin-wave dispersion.

\begin{figure}[h!]
\begin{center}
\includegraphics[width=\columnwidth]{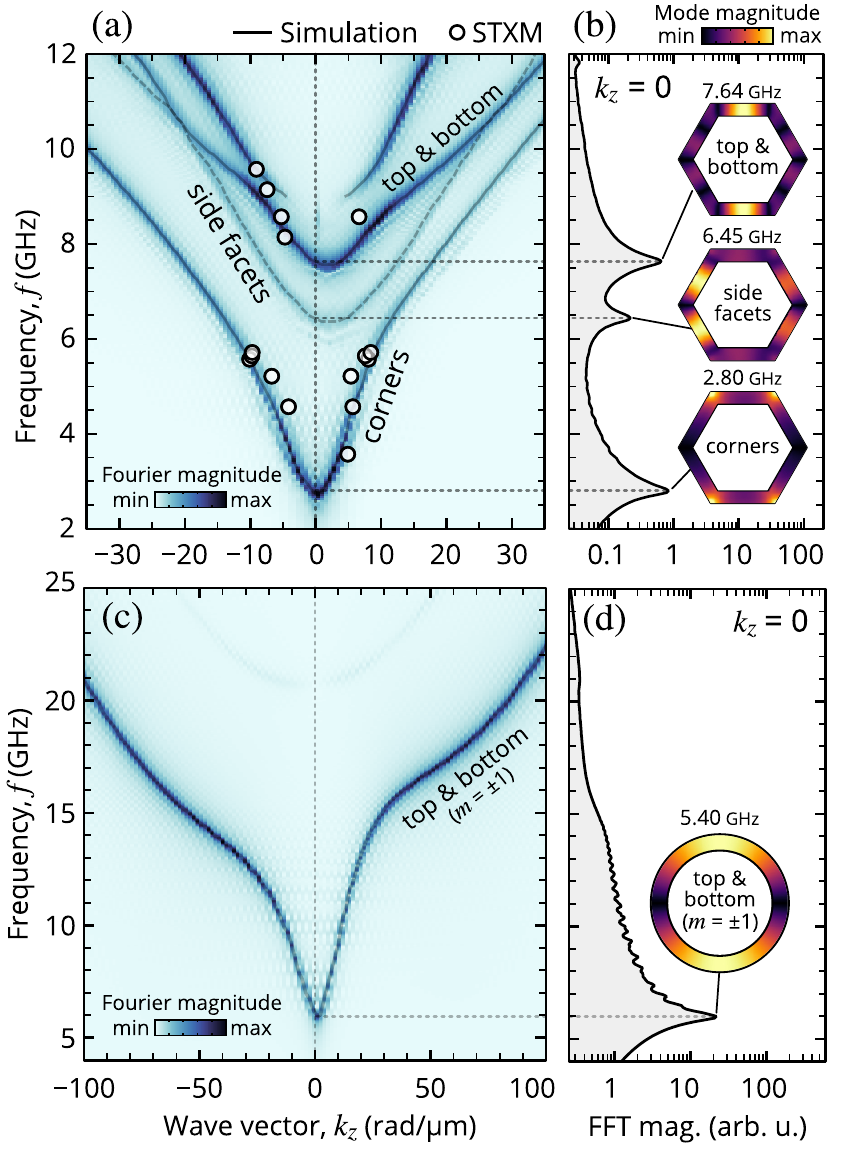}
\caption{\label{fig:fig3}(a) Numerically and experimentally obtained dispersion of the spin waves in a hexagonal permalloy nanotube with \SI{250}{\nano\meter}  outer diameter and \SI{30}{\nano\meter} thickness, excited with a strip-line microwave antenna at the center of the tube. The spectra were obtained by micromagnetic simulations (using $\textrm{MuMax}^3$~\cite{Vansteenkiste14}), shown as heat map, and TR-STXM, shown as dots on top of the map, respectively. (b) The branches of the dispersion are categorized by their cross-sectional mode profiles at $k_z=0$ which show waves propagating at the corners of the hexagonal cross-section (\SI{2.80}{\giga\hertz}), along the side facets (\SI{6.45}{\giga\hertz}) and along the top and bottom facets (\SI{7.64}{\giga\hertz}). The dispersion is found to be asymmetric with respect to inversion of the wave vector $k_z$ for all branches. For comparison, (c)  shows the numerically obtained asymmetric spin-wave dispersion relation of a round vortex-state nanotube with equal dimensions, material parameters and excitation scheme (note the different scales in frequency and wave vector). Panel (d) shows the corresponding power spectrum of the dynamical magnetization at wave number $k_z=0$. The main peak at $f=\SI{5.40}{\giga\hertz}$ corresponds to the dynamics of modes traveling on the top and bottom half of the tube.}
\end{center}
\end{figure}

To obtain the full dispersion we extracted the wave vectors for different excitation frequencies. For this, we average the experimentally measured data along the width direction of the tube (shown on top of the STXM snapshots in Fig.~\ref{fig:fig2}) and fit these curves for individual time frames with decaying sinusoidal functions along the long axis of the tube (propagation direction). This method was chosen because only a few wavelengths are observed in the measurements, and a Fourier analysis to obtain the wave vectors at a given frequency was not conclusive. In order to interpret the experimental results 
we additionally calculate the dispersion using micromagnetic simulations. The antenna geometry and material parameters including the easy-plane anisotropy, are the same as in the experiments. In the simulations we use a tube length of \SI{10}{\micro\meter}. After having excited all frequencies up to a cut-off frequency of $f_{c}=\omega_c/2\pi=\SI{50}{\giga\hertz}$ homogeneously using a $\sin(\omega_c t)/(\omega_c t)$ pulse and letting the magnetization dynamics evolve for \SI{25}{\nano\second}, the wave-vector dependent frequencies are obtained using a fast Fourier transform (FFT) in time and along the tube's main ($z$) axis. The resulting dispersion is shown in Fig.~\figref{fig:fig3}{a}. The experimental values are added as dots on the numerical dispersion in Fig.~\figref{fig:fig3}{a} and show a very good agreement with the simulation.

For comparison, in Fig.~\figref{fig:fig3}{c}, we show the numerically obtained dispersion for a round tube with the same antenna geometry, material parameters and more importantly, equal excitation conditions. As seen in the dispersion calculated with our finite element code TetraMag~\cite{Kakay10}, the field pulse in this geometry couples to the modes propagating on the top- and bottom sides of the round tube (which may be referred to as modes with an azimuthal index of $m=\pm 1$~\cite{Otalora17}). We immediately observe that the mode spectrum in the hexagonal tube is much richer than for the round tube. It can be seen that, using the single wire antenna, a number of dispersion branches can be excited compared to the single mode in round tubes, although with different efficiency, which is reflected by the FFT intensities.


To classify the different branches of the dispersion, we performed an additional set of micromagnetic simulations. This time we apply a sinc pulse in time, using a homogeneous excitation field profile to excite mainly the modes with wave vector $k_z=0$; the corresponding Fourier spectrum is plotted in Fig.~\figref{fig:fig3}{b}. Performing a windowed inverse Fourier transform at the frequencies of the three lowest peaks reveals the spatial mode profiles. Cross-sections of these profiles at the center of the tube are shown as insets in Fig.~\figref{fig:fig3}{b}. We find that the peak lowest in frequency at \SI{2.80}{\giga\hertz} corresponds to spin waves which are localized in the corners of the hexagonal tube. In accordance to this, recall, that \textit{e.g.} the mode excited at $f=\SI{5.571}{\giga\hertz}$ (shown in Fig.~\figref{fig:fig2}{a}) exhibited the same localization as it belongs to the this particular branch of the dispersion. Above in frequency, the peak at \SI{6.45}{\giga\hertz} in Fig.~\figref{fig:fig3}{b} corresponds to the modes localized on the side facets of the tube. This explains why the second branch in the dispersion is much weaker in intensity, because the antenna at the bottom of the tube is not able to efficiently couple to these modes. As a consequence, in the experiments we can neglect any contribution of the side-facet modes. Finally, the third branch in the dispersion, in Fig.~\figref{fig:fig3}{b} represented by the peak at \SI{7.64}{\giga\hertz} corresponds to modes propagating on the top-and-bottom facets of the hexagonal tube. The mode snapshots presented earlier in Fig.~\figref{fig:fig2}{b,c} belong to this dispersion branch. These modes are most similar to those in the Damon-Eshbach geometry~\cite{Damon1961} in thin films. Additionally, there is a fourth branch which does not have any intensity at $k_z=0$ and is, therefore, not seen in the power spectrum in Fig.~\figref{fig:fig3}{b} but only in the full dispersion Fig.~\figref{fig:fig3}{a}. Indeed, these modes correspond to the azimuthal ($m=\pm 1$) modes known from round nanotubes (Fig.~\figref{fig:fig3}{d}). The different branches of the dispersion exhibit multiple (avoided and unavoided) level crossings. Note, that a thorough discussion of the modal spectrum presented here would go beyond the scope of this Letter and will be published elsewhere.

\begin{figure}[h!]
\begin{center}
\includegraphics[width=\columnwidth]{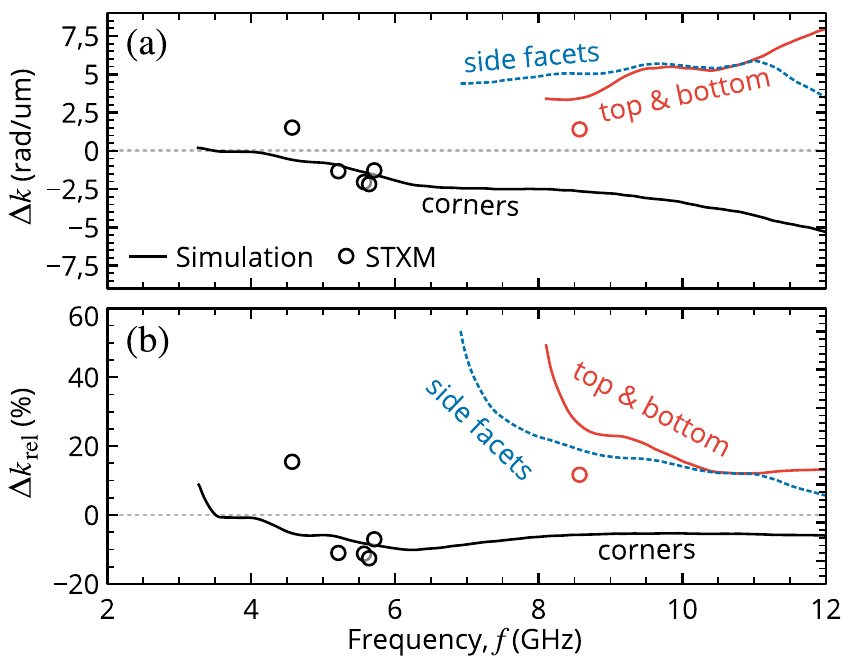}
\caption{\label{fig:fig4} Absolute and relative wave-vector asymmetry, $\Delta k = k_+-k_-$ and $\Delta k_\mathrm{rel} = \Delta k /(k_+ +k_-)$, obtained by micromagnetic simulation and TR-STXM using the dispersion data presented in Fig.~\figref{fig:fig3}{a}.}
\end{center}
\end{figure}

Finally, we discuss the asymmetry of the dispersion. Although from the snapshots obtained by micromagnetic simulations it can clearly be seen that the counter-propagating waves at the same frequency inherit a different wave number. This might not be immediately obvious from the experimental data.
In order to quantify the left-right asymmetry of the spin-wave propagation, we define the absolute difference in wave vector $\Delta k$ for counter-propagating waves at the same frequency, $\Delta k(f) = k_+(f)-k_-(f)$, as the asymmetry of the dispersion. From the dispersion relation shown in Fig.~\figref{fig:fig3}{a} one can already conclude that the asymmetry is present for all branches of the dispersion. In Fig.~\ref{fig:fig4} the absolute asymmetry for the corner as well as the top-bottom facet modes obtained from simulations (solid lines) and from experiments (hollow dots) is plotted together with the relative asymmetry $\Delta k_\mathrm{rel}=\Delta k/(k_++k_-)$. Unfortunately, from the experiments we were not always able to obtain satisfactory fits for both propagation directions for all the measured frequencies, thus $\Delta k$ could not always be extracted. However, as seen in Fig.~\ref{fig:fig4}, the experimental results are still in a qualitative agreement with the micromagnetic simulations.

The overall sign of the asymmetry $\Delta k$ depends on the dispersion branch. This feature is also present for round nanotubes for which the sign can also change for branches with different azimuthal indices $m$. Notably, not only the corner modes propagating in the highly curved regions of the hexagonal nanotube are strongly influenced by the curvature-induced magnetochiral effects. There is also a significant influence on the modes that propagate localized to the facets which are, for themselves, flat objects. But still they constitute to a net curvature if combined into a tube. Next to the overall dispersion asymmetry, one can clearly see from Fig.~\figref{fig:fig3}{a} that the saddle points or bottoms of the dispersion branches are already fairly shifted with respect to $k_z=0$ for the side facet and the top-bottom-facet modes. In case of spin-wave transport in the Damon-Eshbach configuration, as in the case of the current manuscript, this is an indication that an analytical expression for the dispersion contains a term which is, in first order approximation, linear in the wave number $k_z$. This is common for the spin waves in systems with a magneto-chiral interaction, \textit{e.g.} thin-film samples with intrinsic Dzyaloshinskii-Moriya interaction~\cite{Cortes-Ortuno13}, and, as recently shown, round nanotubes in the vortex state~\cite{Otalora16,Otalora17}.

Let us note, that the antiparallel alignment of the equilibrium magnetization in opposite facets would alone lead to an asymmetry, namely to a linear shift of the dispersion in the small $k$ limit \cite{Gallardo2019}. This effect contributes as an approximately hyperbolic decay in the relative wave-vector asymmetry with respect to the excitation frequency $f$ (see Supplemental Material). We see in Fig.~\figref{fig:fig4}{b}, that this is not the case for the hexagonal nanotube meaning that the asymmetry is strongly influenced by the curvature. Discussing the relation between this long-range dipolar coupling of the two opposite facets and the curvature-related effects would go beyond the scope of this Letter. However, both effects have their origin in the nontrivial spatial distribution of the dynamic dipolar fields in the tube.

In conclusion, we showed spin-wave propagation in three-dimensional permalloy nanotubes of hexagonal cross section using TR-STXM measurements. We demonstrate experimentally, that the curvature-induced magneto-chiral effects originating from the dipole-dipole interaction are directly observable in agreement with numerical and analytical predictions. We find that the dispersion relation is asymmetric regarding the wave vector, therefore spin waves of the same frequency propagating into opposite directions have different wavelengths. Moreover, the magneto-chiral effect and therefore the asymmetric spin-wave transport originating from the dipole-dipole interaction is present for all modes, even for those localized to the flat facets. Since hexagonal nanotubes can be manufactured with a very high quality and precision, we believe, that they are a perfect model system for the emerging field of spin-wave transport in curvilinear geometries. Therewith, hexagonal tubes provide an opportunity to unveil novel functionalities, such as unidirectional spin-wave transport in non-reciprocal wave guides induced by the curvature.

The experiments were mainly performed at the MAXYMUS endstation of BESSY II at Helmholtz-Zentrum Berlin, Germany. We thank HZB for the allocation of synchrotron radiation beam time. Some experiments were performed at the PolLux endstation of the Swiss Light Source. We acknowledge the Paul Scherrer Institut, Villigen PSI, Switzerland for provision of synchrotron radiation beamtime. The PolLux end station was financed by the German Ministerium für Bildung und Forschung (BMBF) through contracts 05K16WED and 05K19WE2. Financial support by the Deutsche Forschungsgemeinschaft within the program KA 5069/1-1 is gratefully acknowledged. We also gratefully acknowledge financial support by the Fondecyt Iniciacion grant number 1119018.

\input{main.bbl}
\bibliographystyle{apsrev4-1}


\end{document}

%% file: main.bbl
%